\documentclass[page-classic]{epl2} 

\usepackage{amsmath}
\usepackage{graphicx} 

\def\beq{\begin{equation}}
\def\eeq#1{\label{#1}\end{equation}}
\def\beqa{\begin{eqnarray}}
\def\eeqa#1{\label{#1}\end{eqnarray}}

\title{Critical fluctuations of noisy period-doubling maps}

\author{Andrew E.~Noble\inst{1,2} \and Saba Karimeddiny\inst{2} \and Alan Hastings\inst{1} \\ \and Jonathan Machta\inst{1,3}}
\shortauthor{Andrew E.~Noble \etal}

\institute{                    
  \inst{1} Department of Environmental Science and Policy, University of California, Davis, CA 95616, USA \\
  \inst{2} Department of Physics, University of Massachusetts, Amherst, Massachusetts 01003, USA \\
  \inst{3} Santa Fe Institute, 1399 Hyde Park Road, Santa Fe, NM 87501, USA
}

\abstract{We extend the theory of quasipotentials in dynamical systems by calculating, within a broad class of period-doubling maps, an exact potential for the critical fluctuations of pitchfork bifurcations in the weak noise limit.  These far-from-equilibrium fluctuations are described by finite-size mean field theory, placing their static properties in the same universality class as the Ising model on a complete graph.  We demonstrate that the effective system size of noisy period-doubling bifurcations exhibits universal scaling behavior along period-doubling routes to chaos.}

\begin{document}

\maketitle

Period-doubling bifurcations have been observed in a wide variety of natural systems spanning many areas of science~\cite{Ott:2002wz}.  Univariate, discrete-time maps are the simplest dynamical systems to exhibit a period-doubling route to chaos~\cite{Feigenbaum:1979up,Feigenbaum:1978th}.  Applications range from the population dynamics of species with non-overlapping generations~\cite{May:1974uz,Hastings:1993p13061} to the oscillations of rf-driven Josephson junctions~\cite{Huberman:1980hs,Kautz:1981uh}.  The impact of noise on period-doubling maps has been extensively studied in both ecology~\cite{Schaffer:1986vo,Vilar:1998vv, Melbourne:2008p9904,Dobramysl:2013uj} and physics~\cite{MayerKress:1981tl, Geisel:1982uo, Baras:1982vy, VandenBroeck:1982vy, Svensmark:1987vg, Svensmark:1990uq, Weiss:1987vp,Neiman:1994th,Neiman:1997td,Wang:1999tb,Omberg:2000wv,Kapustina:2002wb,McKane:2005p8516,  Giardina:2006vn, Sole:2006uv, Sole:2011us, Challenger:2013tw, Challenger:2014ep, ParraRojas:2014wb}, including the universal scaling of the Lyapunov exponent along period-doubling routes to chaos~\cite{Crutchfield:1980ud,Huberman:1980wc,Crutchfield:1981ts,Shraiman:1981dx,Crutchfield:1982ub,Feigenbaum:1982tj}.  Connections between such noisy dynamical systems and the universality classes of equilibrium statistical physics has been a subject of great fascination~\cite{Hohenberg:1977ym,Grinstein:1985ei,Crutchfield:1990wx,Tauber:1992wi,VandenBroeck:1994iy,Marcq:1996ec,VandenBroeck:1997wd,Egolf:2000we,Sastre:2001ex,Pikovsky:2002kw,Risler:2004ij,Marcq:2006kt,Wood:2006un}.  The theory of quasipotentials, providing a formal link between equilibruim and nonequilibrium physics, has been applied to many systems~\cite{Freidlin:2012wd,Kifer:1988wd}.  In particular, the theory of quasipotentials has been used to estimate escape times from the attractors of noisy period-doubling maps~\cite{Beale:1989ta,Graham:1991p10465,Reimann:1995wk} and to estimate the invariant probability distributions of their strange attractors in the chaotic regime~\cite{Graham:1991p10465,Reimann:1991vz,Hamm:1992tk}.  

Here, we investigate the invariant probability distributions that characterize critical fluctuations in the pitchfork bifurcations of period-doubling maps far from the chaotic threshold.  In the limit of weak noise, we find an exact correspondence between the static behavior of fluctuations at a pitchfork bifurcation and the critical behavior of {\it finite-size} mean field theory~\cite{Brezin:1985vt}.  This correspondence places pitchfork bifurcations in the same universality class as the Ising model on a complete graph~\cite{Ellis:1978wt,Klein:2007uy,Ellis:2008tp,Ellis:2010wi,ColonnaRomano:2014tya}.    Analytical estimates of critical exponents and amplitudes agree well with the results of numerical simulations.  We conclude with evidence of universal scaling behavior in the effective system size of critical fluctuations along period-doubling routes to chaos.  

We start by considering a noisy, one-dimensional map that can be written as
\beq
x_{t+1}\,=\,\mu(r,x_t) + \lambda \sigma(r,x_t) \xi_t,
\eeq{nm}
where the $x_t$ are random variables, $r$ is the control parameter, $\mu(r,x_t)$ is a deterministic period-doubling map, $\lambda>0$ is an overall constant factor setting the noise level, $\sigma(r,x_t)$ is a nonnegative function, and the $\xi_t$ are independent, identically distributed standard normal random variables.  We restrict our attention to fluctuations arising from the effects of weak noise on supercritical pitchfork bifurcations.  A pitchfork bifurcation in a period-doubling sequence corresponds to a loss in stability of the steady state fixed points, $x^*$, with $(r_c, x^*_c)$ being the critical point where~\cite{Ott:2002wz,Kuznetsov:2004,Golubitsky:2013ud}
\beqa
\mu(r_c,x^*_c)&=&x^*_c, \nonumber \\
\partial_x\mu(r_c,x^*_c)&=&-1.
\eeqa{breq}
Note that $x^*$ is the stable fixed point for $r<r_c$, and we define it to be the unstable fixed point for $r>r_c$. The first bifurcation breaks a time-translation symmetry:  below $r_c$, the stable attractor is a steady-state fixed point, and trajectories on the attractor are invariant to all time-translations; above $r_c$, the stable attractor is a two-cycle and corresponding trajectories are only invariant to translations by an even-number of time steps.  We will investigate the critical fluctuations that emerge from this ${\cal Z}_2$ symmetry breaking in the limit of weak noise ($\lambda\rightarrow 0$).   Our results rely on the additional assumptions that $\mu(r,x_t)$ is a $C^3$ map~\cite{Singer:1978vg}, i.e.~three-times differentiable with respect to $x_t$, and that the stable basin of attraction of $\mu(r,x_t)$ is bounded on some finite interval.

Noisy maps in the form of Eq.~\ref{nm} have been applied to the study of stochastic fluctuations in ecological population densities~\cite{Schaffer:1986vo, Melbourne:2008p9904}.  If $\mu(r,x_t)$$=$$rx_t(1-x_t)$ and $\sigma(r,x_t)$$=$$1$, then Eq.~\ref{nm} is the logistic map with extrinsic additive noise.  Non-constant choices for $\sigma(r,x_t)$ model intrinsic noise or a combination of intrinsic and extrinsic noise.  When applied to population dynamics, common choices might be $\sigma(r,x_t)$$=$$\mu(r,x_t)$, modeling multiplicative environmental noise, or $\sigma(r,x_t)$$=$$\sqrt{\mu(r,x_t)}$, modeling demographic noise.  

Over long enough time scales, a noisy trajectory, $x_t$, will depart from the stable basin of attraction of $\mu(r,x_t)$.  However, for small enough $\lambda$ and for initial conditions on the interval bounding the stable attractor of $\mu(r,x_t)$, a metastable probability distribution for $x_t$ will persist over ``intermediate" time periods that are much longer than the duration of initial transients but exponentially shorter than the characteristic time to depart from the stable attractor.  

\begin{figure}
\hskip-0.4cm\centerline{\includegraphics[width=80mm]{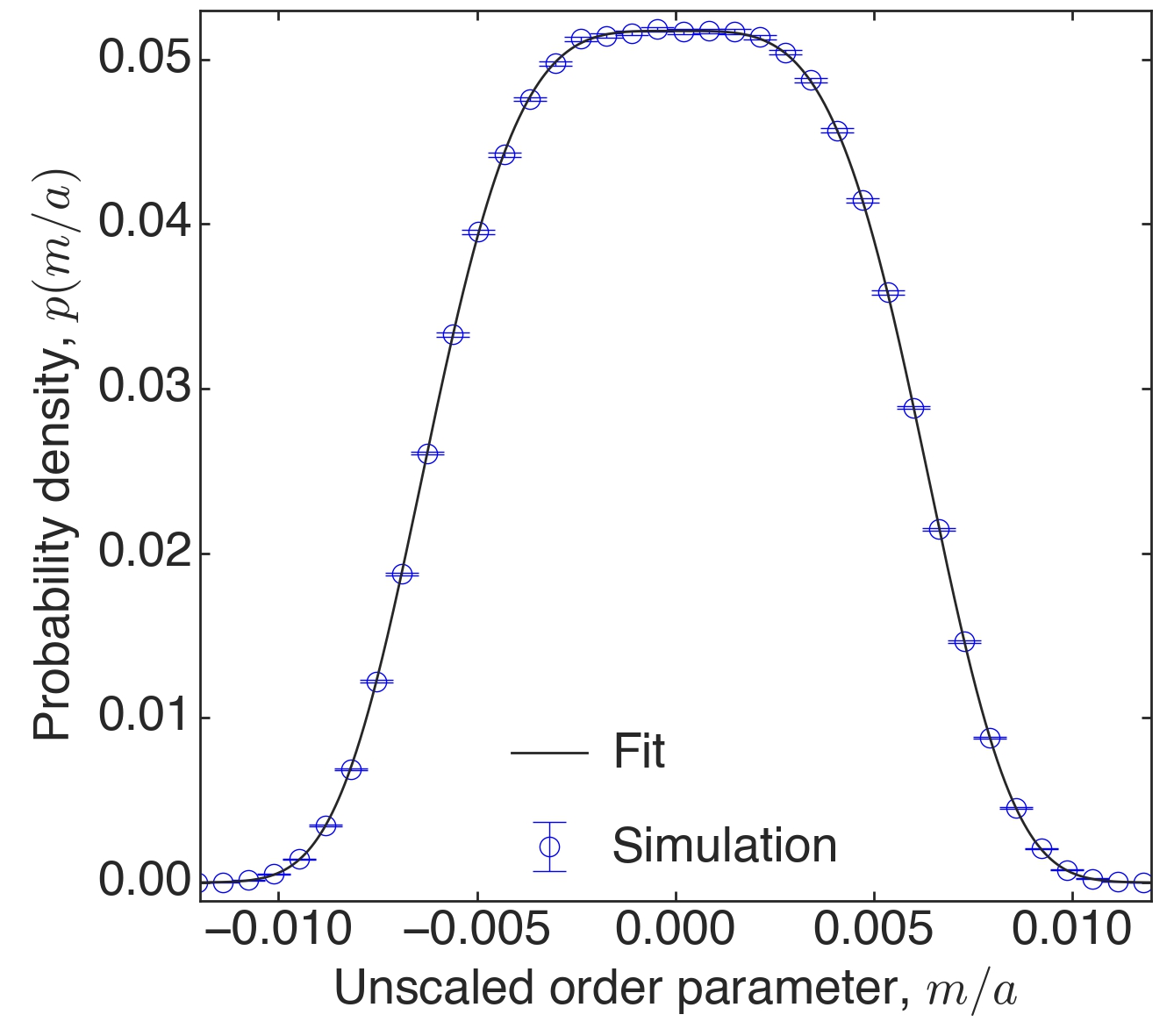}}%
\vskip-0.3cm
\caption{(Color online) Histogram of the unscaled order parameter, $m / a$, at a noisy supercritical pitchfork bifurcation.  Blue points and error bars are estimated from numerical simulations of Eq.~\ref{nm} at the first period-doubling transition of a logistic map, $\mu(r,x)$$=$$r x(1-x)$ and $r$$=$$r_c$$=$$3$, with additive noise, $\lambda$$=$$10^{-4}$ and $\sigma(r, x)$$=$$1$.  Similar results hold for all $\lambda<10^{-2}$.  The black line is a fit of a pure quartic distribution to the simulation results (chi-square per d.o.f.~1.3).  Fluctuations in the pitchfork order parameter described by a quartic distribution suggest a correspondence between the critical behavior of noisy pitchfork bifurcations and finite-size mean field theory.}
\label{fig1}
\end{figure}

This separation of time scales allows us to quantify the critical fluctuations of pitchfork bifurcations.  The metastable distribution of the map variable, $x_t$, is approximately stationary at any given intermediate time, $T$.  We define the pitchfork order parameter, $m$, as 
\beq
m\,\equiv\, \frac{a}{2} (x_{T+1}  - x_T),
\eeq{op}
where $a$ is a model-dependent scale factor that maps the pitchfork order parameter onto the Ising mean-field magnetization.  To obtain numerical estimates of the order parameter, we use ergodicity~\cite{Reimann:1991vz} and sample statistics along a single trajectory of Eq.~\ref{nm} at intermediate time scales.

Fig.~\ref{fig1} is a histogram of the unscaled order parameter, $m / a$, for a logistic map with weak additive noise.  The distribution of the order parameter is highly symmetric and can be well-approximated by a pure quartic.  These numerical results motivate us to seek an analytical correspondence between the critical fluctuations of pitchfork bifurcations and those of finite-size mean field theory for the Ising model.  The latter are described by the distribution 
\beq
p_{\rm MFT}(m)\,\propto\,\exp\left(-Nf(m)\right),
\eeq{cw}
where $N$ is the number of spins and the free energy (the large-deviation potential) at reduced temperature $\tilde{t}$ is
\beq
f(m)\,=\,\frac{1}{2} \tilde{t} m^2+\frac{1}{4!} m^4,
\eeq{fe}
to leading order in $N$.  
	
The first step is to calculate the metastable distribution, $h(x)$, of the map variable, $x$, near the first period-doubling pitchfork bifurcation in the weak noise limit.  We re-write the noisy map of Eq.~\ref{nm} in integral form and obtain a self-consistent equation for $h(x)$
\beq
h(x)\,=\,\int dx^\prime h(x^\prime)k(x^\prime,x),
\eeq{hxeq}
where the kernel describing the behavior of the noisy map is given by
\beq
k(x^\prime,x)\,=\,\frac{1}{\sqrt{2\pi}\lambda\sigma(r,x^\prime)}\exp\left(-\frac{\left(x-\mu(r,x^\prime)\right)^2}{2 \lambda^2\sigma(r,x^\prime)^2}\right).
\eeq{k}
For weak noise, $h(x)$ will be sharply peaked near $x^*$, and we can take the limits of integration in Eq.~\ref{hxeq} from $-\infty$ to $\infty$.  Near the critical point, $x^*$ is approximately the same as $x^*_c$ with the leading order correction given by
\beq
x^*-x_c^*\,=\,-\frac{\partial_r \mu(r_c, x_c^*)}{2+\partial_r\partial_x \mu(r_c, x_c^*)}\left(r_c - r\right).
\eeq{xslam}
We define $\tilde{h}(x-x^*)=h(x)$ and make an exponential ansatz for $\tilde{h}$ given by
\beq
\tilde{h}(z)\,\propto\,\exp\left(-\frac{1}{2\lambda^2\sigma(r_c, x_c^*)}\left[b_{2,1} z^2 \delta + b_{3,1} z^3 \delta + b_{4,0} z^4 + b_{5,0} z^5 +\cdots\right]-\left[a_{1, 0} z+\cdots\right] \right),
\eeq{hx}
where $\delta=r_c - r$ and the dots indicate higher order terms.  In the limit of small $\lambda$, we apply Laplace's method, and after some algebra, find a self-consistent solution to Eq.~\ref{hxeq} where
\beqa
a_{1, 0}&=& \frac{1}{2}\partial_{xx}\mu(r_c,x^*_c), \nonumber \\
b_{2,1}&=& -T\mu(r_c, x^*_c), \nonumber \\
b_{3,1}&=& \frac{1}{2}T\mu(r_c, x^*_c)\partial_{xx}\mu(r_c,x^*_c), \nonumber \\
b_{4,0} &=& -\frac{1}{6}S\mu(r_c, x^*_c), \nonumber \\
b_{5,0} &=& \frac{1}{6}S\mu(r_c, x^*_c)\partial_{xx}\mu(r_c,x^*_c). 
\eeqa{ab}
The expressions in Eq.~\ref{ab} depend on the critical values of the Schwarzian derivative~\cite{Singer:1978vg}
\beq
S\mu(r_c, x^*_c)\,\equiv\,-\left(\partial_{xxx}\mu(r_c,x^*_c) + \frac{3}{2}\left(\partial_{xx}\mu(r_c,x^*_c)\right)^2\right),
\eeq{s}
and the ``T-derivative"
\beq
T\mu(r_c, x^*_c)\,\equiv\, -\left.\partial_r\partial_x \mu(r,\mu(r,x))\right\vert_{r=r_c,x=x^*_c}.
\eeq{treq}
The Schwarzian derivative arises from a second-order stability analysis of the normal form of a pitchfork bifurcation and is always negative for the supercritical pitchfork bifurcations of $C^3$ maps~\cite{Singer:1978vg}.  The T-derivative arises in our analysis due to the expansion in $\delta$ and is always negative for pitchfork bifurcations with a symmetric phase below $r_c$ and a broken symmetry above, as can be seen from a calculation using the normal form~\cite{Kuznetsov:2004}. 

Note that the self-consistent solution for $\tilde{h}$, as given in Eq.~\ref{hx}, is not symmetric under $z\rightarrow -z$ due to the odd terms in $z$, whereas the order parameter is expected to be symmetric under $m\rightarrow -m$, at least to leading order.  Given the definition of the order parameter in Eq.~\ref{op}, we have
\beq
p(m)\,=\,\int\int~dx^\prime~dx~\delta\left(m - \frac{a}{2}[x - x^\prime]\right)h(x^\prime)k(x^\prime, x).
\eeq{pm}
Upon calculating the integral, we find that $p(m)=p_{\rm MFT}(m)$, as defined in Eqs.~\ref{cw} and \ref{fe}, to leading order in $m$ and $N$, with order parameter rescaling factor 
\beq
a\,=\,\sqrt{\frac{2S\mu(r_c,x^*_c)}{T\mu(r_c,x^*_c)}},
\eeq{aval} 
effective reduced temperature 
\beq
\tilde{t}\,=\,r_c-r,
\eeq{ttval} 
and effective system size 
\beq
N\,=\,\frac{1}{2\lambda^2\sigma^2(r_c,x^*_c)}\left[\frac{\left(T\mu(r_c,x^*_c)\right)^2}{-S\mu(r_c,x^*_c)}\right].
\eeq{Nval} 
Consistent with previous calculations of quasipotentials in noisy dynamical systems~\cite{Freidlin:2012wd,Kifer:1988wd,Graham:1991p10465,Beale:1989ta,Hamm:1992tk}, the expression for the effective system size parameter, $N$ in the exponent of Eq.~\ref{cw}, is inversely proportional to the variance of the noise, $\lambda^2\sigma^2(r,x^*_c)$.  Note that the static behavior of critical fluctuations at a pitchfork bifurcation is entirely determined by $N$.  In applications to ecological systems with demographic noise due to the well-mixed interactions of a finite, rather than an infinite, number of individuals~\cite{Melbourne:2008p9904}, $N$ is directly proportional to population size.  Our results quantify the scale of fluctuations for finite-size populations undergoing a period-doubling transition.

\begin{figure}[b!]
\centerline{
\hskip-0.2cm\includegraphics[width=80mm]{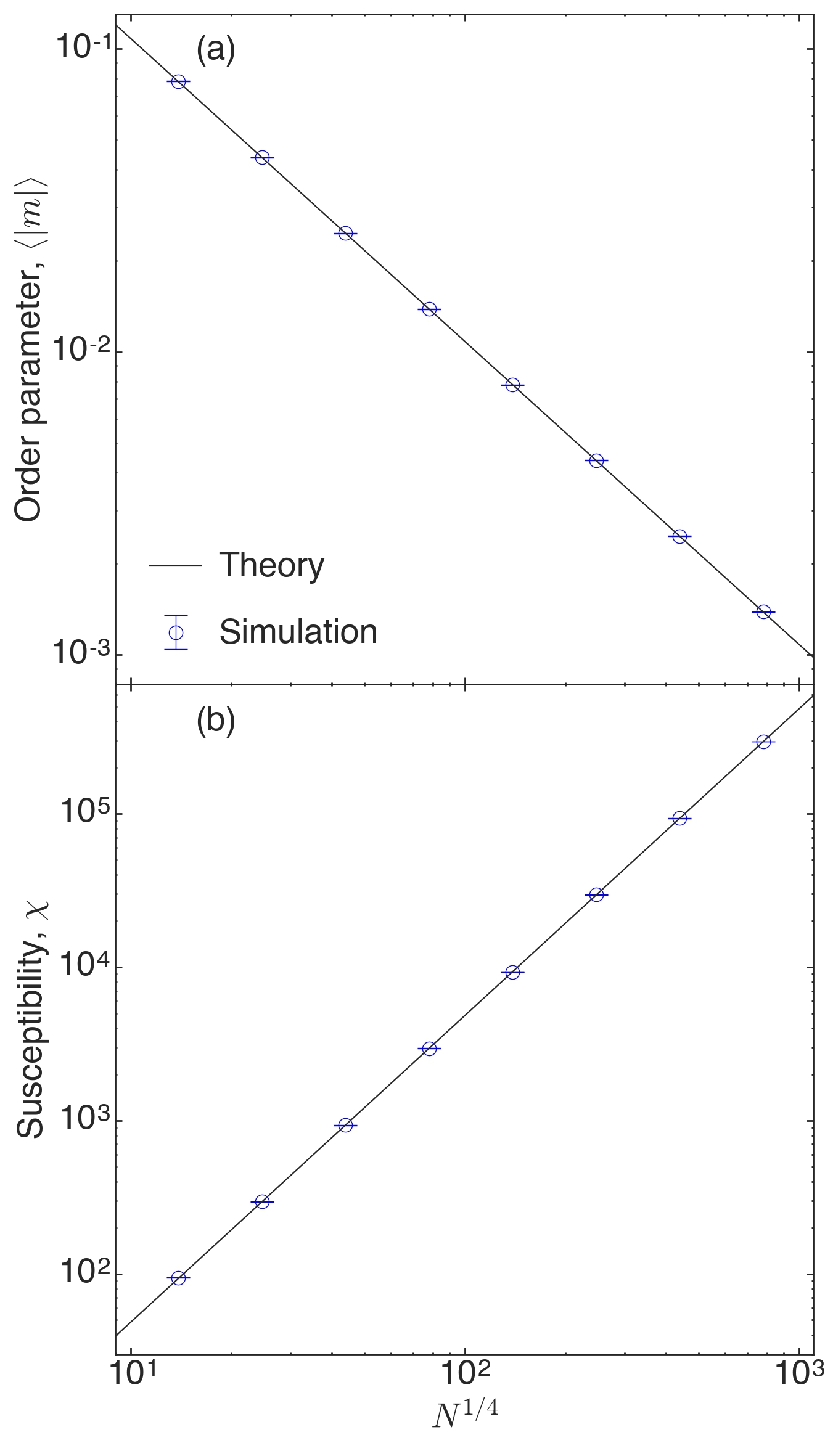}}
\vskip-0.3cm
\caption{Finite-size scaling of (a) the absolute value of the pitchfork order parameter, $\langle\left\vert m \right\vert \rangle$, and (b) the susceptibility, $\chi$.  Numerical estimates (symbols and error bars) are based on samples of the order parameter along Monte Carlo trajectories for the the first period-doubling pitchfork bifurcation of a logistic map with weak additive noise, i.e.~Eq.~\ref{nm} with $\mu(r,x)$$=$$r x(1-x)$, $\sigma(r, x)$$=$$1$, and $r$$=$$r_c$$=$$3$.  Numerical estimates of $\langle\left\vert m \right\vert \rangle$ and $\chi$ are consistent (chi-square per d.o.f.~of $0.35$ and $2.5$, respectively) with the predictions of finite-size mean field theory (Eqs.~\ref{fssp:m} and \ref{fssp:chi}, respectively) and $N$ as defined by Eq.~\ref{Nval}.  
\label{fig2}}
\end{figure}

For pitchfork bifurcations, the leading odd correction to the potential in Eq.~\ref{fe} is generally ${\cal O}(m^7)$.  However, an important quadratic map in ecology is the Ricker map~\cite{Ricker:1954wv}, where
\beq
\mu(r, x_t)\,=\,x_t\exp(r(1-x_t)),
\eeq{ricker}
and this map is symmetric about the first period-doubling bifurcation at $(r_c, x^*_c) =  (2, 1)$, such that the second derivative $\partial_{xx}\mu(r_c,x^*_c)$ vanishes and the odd correction to the potential at ${\cal O}(m^7)$ is zero.  We speculate that all odd corrections vanish in the potential describing critical fluctuations near the first period-doubling bifurcation of the Ricker map.

\begin{figure}[t!]
\centerline{
\hskip-0.08cm\includegraphics[width=80mm]{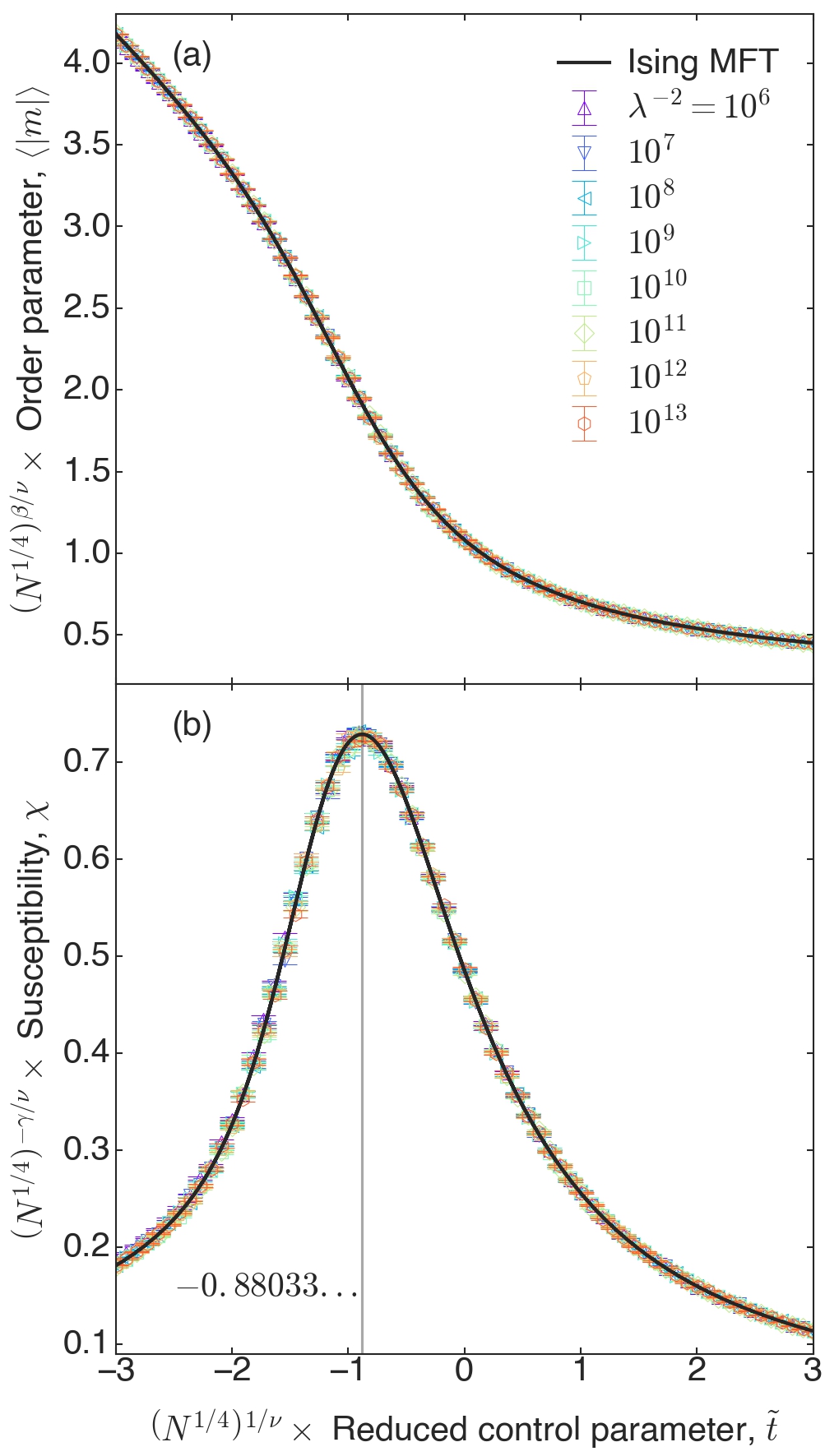}}
\vskip-0.3cm
\caption{Scaling collapse of (a) the absolute value of the pitchfork order parameter, $\langle\left\vert m \right\vert \rangle$ and (b) the susceptibility, $\chi$.  Numerical estimates (symbols and error bars) are based on samples of the order parameter along Monte Carlo trajectories for the the first period-doubling pitchfork bifurcation of a logistic map with weak additive noise, i.e.~Eq.~\ref{nm} with $\mu(r,x)$$=$$r x(1-x)$, $\sigma(r, x)$$=$$1$, and $r$$=$$r_c$$=$$3$.  Numerical estimates of $\langle\left\vert m \right\vert \rangle$ and $\chi$ are consistent with the predictions of finite-size mean field theory (black lines).  In panel (b), the vertical grey line marks the peak value of scaled susceptibility, which defines the finite-size critical point, $r_{c}(N)$ (Eq.~\ref{rcN}).
\label{fig3}}
\end{figure}

\begin{figure}[th!]
\centerline{
\includegraphics[width=80mm]{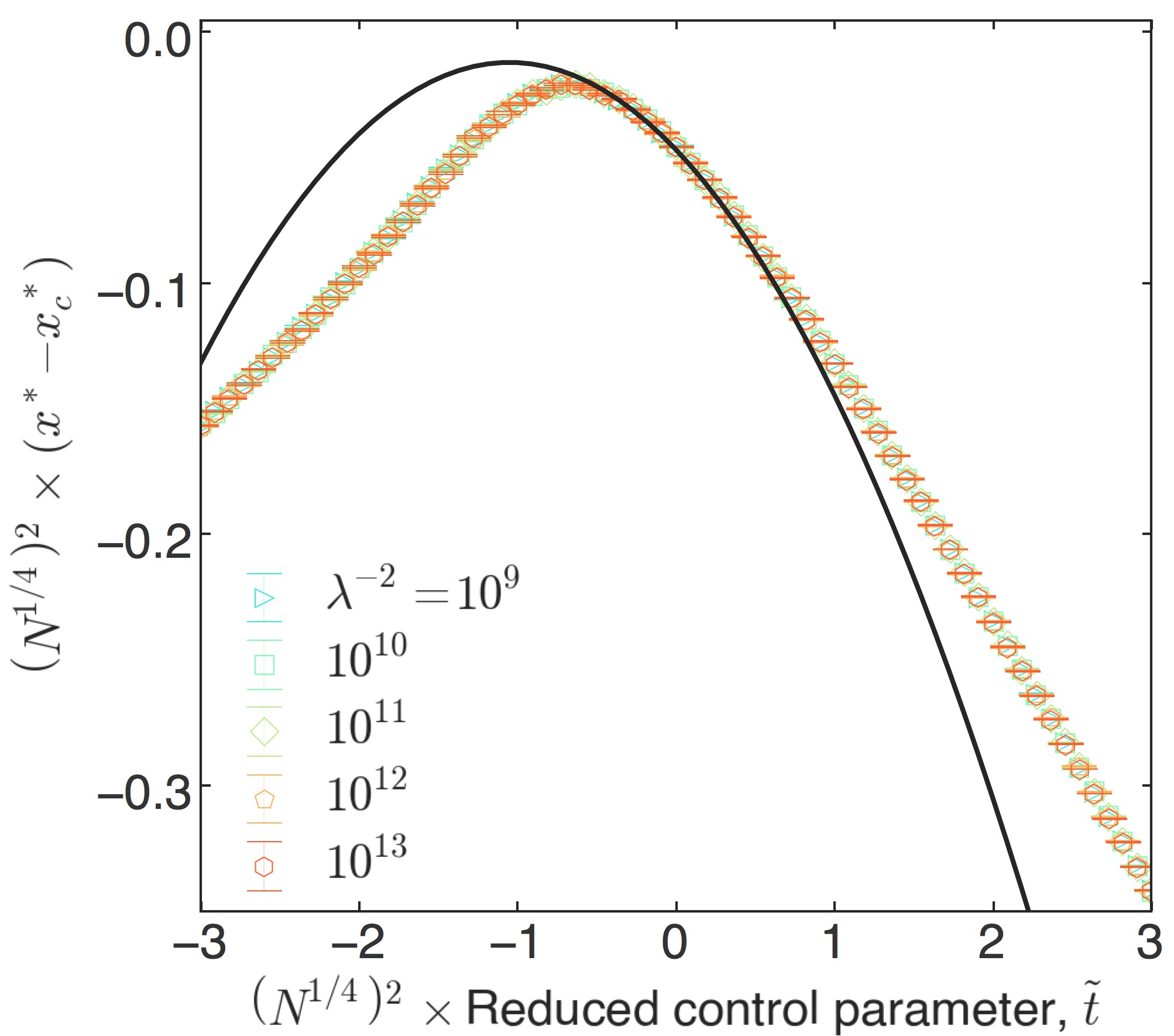}}
\vskip-0.3cm
\caption{Scaling collapse of finite-size corrections to $x^*$.  Numerical estimates (symbols and error bars) are based on sampling values of $x^*$ along Monte Carlo trajectories for the the first period-doubling pitchfork bifurcation of a logistic map with weak additive noise, i.e.~Eq.~\ref{nm} with $\mu(r,x)$$=$$r x(1-x)$, $\sigma(r, x)$$=$$1$, $r$$=$$r_c$$=$$3$, and $x^*_c$$=$$2/3$.  Those numerical estimates are consistent with our theoretical estimate (the first moment of the distribution in Eq.~\ref{hx}, black line) near the critical point, $r=r_c$. 
\label{fig4}}
\end{figure}

We compare mean-field predictions of critical exponents and amplitudes to Monte Carlo simulations~\cite{Binder:1986wu, Janke:1996tm, Janke:2002uc} of the first period-doubling bifurcation in a logistic map with additive noise. Finite-size scaling (Fig.~\ref{fig2}) and scaling collapse (Fig.~\ref{fig3}) are as predicted by finite-size mean field theory.  In the finite-size scaling of the order parameter and the susceptibility, as defined in Eq.~\ref{defs1}, numerical results are consistent with mean-field predictions for critical exponents and amplitudes.  In the scaling collapse of susceptibility, the peak value defines the finite-size critical point
\beq
r_{c}(N)\,=\,r_c + \frac{0.88033\dots}{\sqrt{N}}.
\eeq{rcN}
The scaling collapse of Monte Carlo estimates of $x^*-x^*_c$ is consistent with the prediction of Eq.~\ref{xslam} near $r$$=$$r_{c}$ (Fig.~\ref{fig4}).

We now extend the above results to show that finite-size mean field theory describes the critical fluctuations of all the pitchfork bifurcations of a period-doubling map in the limit of weak noise.  Let $r_n$ (dropping the subscript ``c'' for convenience) denote the $n$th critical point in a period-doubling sequence proceeding from a $q$-cycle of a deterministic map.  On $r_{n-1}$$<$$r$$<$$r_n$, there exist $2^{n-1}q$ stable fixed points of the $n$-times iterated map, $x_{t+1}$$=$$\mu_n(r,x_t)$.  At $r$$=$$r_n$, stability is lost with pitchfork bifurcations occurring at the points $x^*_{n,p}$, where $p$$\in$$\{1,\dots,2^{n-1}q\}$.  With the addition of weak noise, the $n$-times iterated map is, to leading order in $\lambda$
\beq
x_{t+1}\,=\,\mu_n(r,x_t) + \lambda \sigma_n(r,x_t) \xi_t,
\eeq{nmn}
where $\sigma_n(r,x_t)$ is given by a recursion relation~\cite{Shraiman:1981dx} 
\beqa
\hskip-0.5cm\sigma^2_{n+1}(r,x) \,=\, \sigma_n(r, x)^2 \left(\partial_x\mu_n(r, \mu_n(r, x))\right)^2 + \sigma_n(r, \mu_n(r,x))^2.
\eeqa{}

Upon substituting Eq.~\ref{nmn} for Eq.~\ref{nm} and expanding around $(r_n, x^*_{n,p})$ rather than $(r_c, x^*)$, the above results for the critical fluctuations of the first pitchfork bifurcation in a period-doubling map apply to all the pitchfork bifurcations in that map.   Note that the conditions of Eq.~\ref{breq} apply to all supercritical bifurcations, and that if $S\mu(r_c, x^*_c)$$<$$0$, then $S\mu_n(r_n, x^*_{n,p})$$<$$0$ for all $n$ and $p$~\cite{Singer:1978vg}.  

\begin{figure*}[t!]
\centerline{
\hskip-0.4cm\includegraphics[width=140mm]{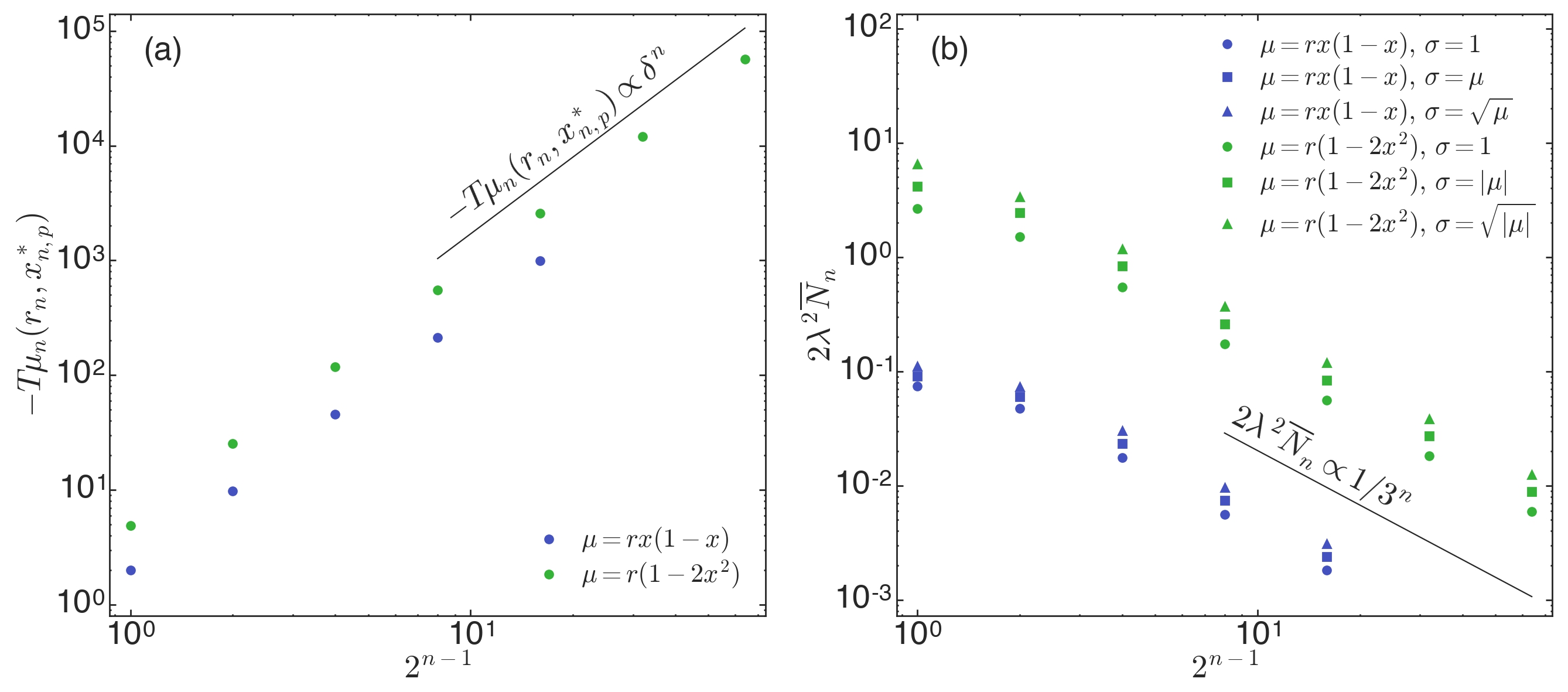}}
\vskip-0.3cm
\caption{(Color online)  Numerical evidence of universality in the scaling behavior of (a) $-T\mu_n(r_n, x^*_{n,p})$, and (b) $2\lambda^2\overline{N}_{n}$.  Panel (a) plots values of $-T\mu_n(r_n, x^*_{n,p})$ along the first period-doubling sequence ($q$$=$$1$) in two different parameterizations of the logistic map.  Panel (b) plots values of $2\lambda^2\overline{N}_{n}$ based on two types of deterministic map, each with three different types of noise.  In both panels, we find evidence of universal scaling behavior in the the approach to the accumulation point, $r$$=$$r_\infty$.
\label{st}}
\end{figure*}

We conclude with a numerical investigation of universal scaling behavior as the first sequence of  period-doubling bifurcations ($q$$=$$1$) approaches the chaotic threshold at  $r$$=$$r_\infty$.  First, in two different parameterizations of the logistic map, we find evidence (Fig.~\ref{st}a) that, for large $n$
\beq
-T\mu_n(r_n, x^*_{n,p}) \,\propto\,(r_\infty - r_n)^{-1} \,\propto\, \delta_F^n,
\eeq{}
where the Feigenbaum constant $\delta_F$$=$$4.6692\dots$ and the value of $-T\mu_n(r_n, x^*_{n,p})$ is independent of $p$.  Second, we investigate scaling in the effective system size of period-doubling bifurcations
\beq
N_{n,p}\,=\,\frac{1}{2\lambda^2\sigma_n^2(r_n, x^*_{n,p})}\left[\frac{\left(T\mu_n(r_n,x^*_{n,p})\right)^2}{-S\mu_n(r_n,x^*_{n,p})}\right].
\eeq{mfmapnp}
For two different deterministic maps and three different types of noise, we calculate the geometric mean of $N_{n,p}$ at fixed $n$
\beq
\overline{N}_{n}\,\equiv\,\left(\prod_{p=1}^{2^{n-1}}N_{n,p}\right)^{1/2^{n-1}},
\eeq{}
and find evidence (Fig.~\ref{st}b) that, roughly
\beq
\overline{N}_{n}\propto \frac{1}{2\lambda^23^n},
\eeq{} 
independent of the details of both the type of deterministic map and the type of noise.  This result is consistent with the intuition that the noise level, $\lambda$, must decrease exponentially with $n$ to maintain the same effective system size, $N_{n,p}$, at each pitchfork bifurcation.  

We have shown that the critical fluctuations of all noisy pitchfork bifurcations are described by finite-size mean field theory for an equilibrium system with $N$ Ising spins.  Our results were obtained for the case of weak Gaussian white noise but are expected to hold more generally for weak noise with a finite second moment and finite correlation time.  This correspondence confirms the breaking of a ${\cal Z}_2$ symmetry at noisy pitchfork bifurcations and introduces $N$ as an effective system size parameter that exhibits universal scaling behavior near the onset of chaos.  In applications to population dynamics with demographic noise, $N$ is directly proportional to the population size.  Further work is needed to obtain a more precise understanding of universal scaling behavior in the critical fluctuations along period-doubling routes to chaos.   

\acknowledgments
We are grateful for useful dialogue with Richard S.~Ellis, William Klein, and Steven H.~Strogatz.  This work is supported by the National Science Foundation under INSPIRE Grant No.~1344187.  S.K.~was partially supported by the National Science Foundation's Soft Matter Research in Theory (SMaRT) Research Experience for Undergraduates.

\section{Appendix A:  Summary of results from finite-size Ising mean field theory}

\renewcommand{\theequation}{A\arabic{equation}}
\setcounter{equation}{0}  

\noindent The existence of finite-size scaling behavior follows from the large-deviation function, Eq.~\ref{fe}, of the order parameter distribution, Eq.~\ref{cw}.  We will compare numerical estimates for the finite-size scaling behavior of noisy pitchfork bifurcations, based on standard Monte Carlo techniques~\cite{Binder:1986wu, Janke:1996tm, Janke:2002uc}, to mean-field predictions for the absolute value of the order parameter, $\langle\left\vert m \right\vert \rangle$, and the susceptibility, $\chi$, given by 
\beq{}
\chi\,\equiv\,N\left(\big\langle\left\vert m \right\vert^2 \big\rangle -\big\langle\left\vert m \right\vert \big\rangle^2\right).
\eeq{defs1}
When evaluated at the critical point, $\tilde{t}=0$, the mean-field observables depend only on  the system size, $N$, namely
\begin{subequations}
\begin{align}
\left.\big\langle\left\vert m \right\vert \big\rangle \right\vert_c&= \frac{1}{\Gamma\left(\frac{1}{4}\right)}\left(\frac{24 \pi^2}{N}\right)^{1/4}, \label{fssp:m} \\
\left.\chi\right\vert_c &=  \frac{4\pi}{\Gamma\left(\frac{1}{4}\right)^2}\left(1-\frac{1}{\sqrt{2}}\right)\sqrt{3N}.\label{fssp:chi} 
\end{align}
\label{fssp}
\end{subequations}
\hskip-0.1cm Near the critical point, expressions for observables can be calculated in closed form from the moments of Eq.~\ref{cw} and obey the finite-size scaling relationships
\beqa
\big\langle\left\vert m \right\vert \big\rangle&=& (N^{1/4})^{-\beta/\nu} G_{m}\left((N^{1/4})^{1/\nu}\tilde{t}\right), \nonumber \\
\chi &=& (N^{1/4})^{\gamma/\nu} G_{\chi}\left((N^{1/4})^{1/\nu}\tilde{t}\right), 
\eeqa{scp}
where $\nu=1/2$, $\beta=1/2$, and $\gamma=1$ are the mean-field critical exponents, and the $G_X$ are universal scaling functions.  We find that Eqs.~\ref{fssp} and \ref{scp}, with $N$ as defined by Eq.~\ref{Nval} or \ref{mfmapnp}, describe the critical fluctuations of pitchfork bifurcations of period-doubling maps in the weak noise limit.

\end{document}